# A FLEXIBLE QUASIOPTICAL INPUT SYSTEM FOR A SUBMILLIMETER MULTI-OBJECT SPECTROMETER

PAUL F. GOLDSMITH & MICHAEL SEIFFERT

Jet Propulsion Laboratory, California Institute of Technology, 4800 Oak Grove Drive, Pasadena CA 91109

**ABSTRACT.** We present a conceptual design for the input optical system for a multi-object spectrometer operating at submillimeter wavelengths. The "Mirror MOS" is based on a sequence of mirrors that enables low-loss propagation of beams from selected positions distributed throughout the focal plane to the spectroscopic receiver inputs. This approach should be useful for observations of sources which have a relatively low density on the sky, for which it is inefficient to use a traditional array receiver with uniformly spaced, relatively closely packed beams. Our concept is based on assigning a patrol region to each of the receivers, which have inputs distributed over the focal plane of the telescope. The input to each receiver can be positioned at any point within this patrol region. This approach, with only 4 reflections, offers very low loss. The Gaussian beam optical system can be designed to produce frequency-independent illumination of the telescope, which is an important advantage for broadband systems such those required for determination of redshifts of submillimeter galaxies.

## 1. INTRODUCTION

There has been considerable development of focal plane arrays for spectroscopy at millimeter wavelengths in recent years. The SEQUOIA array (Erickson *et al*. 1999) operates at 3 mm wavelength and has been upgraded to include 32 receivers, configured as 16 dual linear polarization pixels on the sky. HARP-B (Smith *et al*. 2008) is a 16-element array operating in the 350 GHz (0.85 mm wavelength) on the James Clerk Maxwell Telescope (JCMT). The Supercam 64 element array for 0.85 mm wavelength is under development at the University of Arizona (Groppi *et al*. 2008). Arrays with modest numbers of pixels have been developed for higher frequencies. All of these arrays go to some effort to have their beams on the sky as closely packed as possible. Together with effective methods of telescope scanning and appropriate sophisticated software, these arrays greatly accelerate the rate at which large fields can be mapped, and also result in improved data quality. This results in part from a given position on the sky being observed by multiple receivers, thus minimizing the effect of sensitivity variations.

For observations of extragalactic sources, a different approach is likely necessary. Signal strengths are generally low, so that long integration times will be necessary. This means that there is considerable value in observing multiple sources simultaneously. However, the density of detectable sources on the sky is modest, even for a telescope with relatively large collecting area. Thus, to exploit a focal plane array for spectroscopy, we will likely be observing sources spread across the instantaneous field of view of the telescope, and ideally should be able to match the positions of the array pixels in the focal plane to the positions of the sources of interest which will have been identified and located by less time consuming continuum surveys.

As an example, we consider the challenge of measuring redshifts of submillimeter galaxies (SMGs). These very distant, highly luminous galaxies have so much dust along with young stars that their optical emission (due to the combination of the dust absorption and red shift) is quite weak, and their continuum emission peaks at far-infrared wavelengths. These sources have been intensively studied since their discovery just over a decade ago (Smail, Ivison, & Blain 1997). A review and some recent studies are Blain *et al*. (2002), Laurent *et al*. (2005), Coppin *et al*. (2006), Greve *et al*. (2008), Chapman *et al*. (2008), & Scott *et al*. (2008).

While the number of such sources found to date is modest, large numbers of such interesting sources will likely be found by continuum surveys, carried out with large-format continuum detectors on the JCMT, the Cornell Caltech Atacama Telescope (CCAT, Sebring *et al*. 2008) as well as with other facilities. There are several reasons to consider the 200 to 300 GHz frequency range to determine the redshift of SMGs. The first is that it is a good atmospheric window with relatively large fractional bandwidth, in which high redshift galaxies can be effectively studied through mid-J rotational transitions of carbon monoxide. In order to estimate the number of such sources, we rely on the paper by Blain *et al*. (2000). These authors have modeled the line emission from high redshift galaxies constrained by such observations that are available. They find that in the 210 to 310 GHz range, the emission will be dominated by CO J = 5-4, J = 6-5, and J = 7-6 lines at redshifts between 1 and 2. Their conclusion is that in a frequency interval of 8 GHz, the density of sources at a flux level of $10^{-20}$ $Wm^{-2}$ is 200 $deg^{-2}$. The scaling of source density is not a single power law, but we use the Blain et al. results to infer a density at a flux level of $3x10^{-20}$ $Wm^{-2}$ to be a factor of 6 smaller, or 33 $deg^{-2}$. For this discussion, we assume the frequency range covered to be 80 GHz. Since the distribution with frequency is fairly uniform, our reference density is 330 sources per square degree with a CO line brighter than $3x10^{-20}$ $Wm^{-2}$.

For a specific model we consider the 25m diameter CCAT telescope to have an aperture efficiency at this relatively low frequency equal to 0.6. This results in an effective area of 295 $m^2$ and a sensitivity of 0.11 K/Jy. A source with a line width of 300 $kms^{-1}$ ($\Delta v/c = 10^{-3}$) will, at an input frequency of 250 GHz, have a line width equal to 250 MHz. A flux level of $3x10^{-20}$ $Wm^{-2}$ thus corresponds to a flux density of $1.2x10^{-28}$ $Wm^{-2}Hz^{-1}$ or 0.012 Jy. This will produce an antenna temperature $T_a = 0.0013$ K.

The receiver system for measuring the redshift of SMGs could be a coherent system using either broadband mixers or amplifiers (e.g. Erickson *et al.* 2007). Alternatively, an incoherent system employing bolometers (or other power detectors) together with a spectrograph providing modest resolution (Bradford *et al.* 2008, Nikola *et al.* 2008) could be employed. In the following discussion we assume a coherent system, which offers the possibility of resolving the line profile of a SMG if the signal to noise is sufficient, and which will be only modestly less sensitive than the incoherent system if the rapid recent improvements in high frequency HEMT MMIC amplifiers continues (e.g. Gaier *et al.* 2007; Samoska *et al.* 2008) . To estimate the time required to detect this signal, we assume that we have a dual polarization system with simple switching scheme, so that the rms antenna temperature is given by $\Delta T = \sqrt{2}T_s/\sqrt{(\Delta f\ t)}$, where $T_s$ is the system temperature, $\Delta f$ is the bandwidth, and t is the integration time. For rms noise equal to s times the expected antenna temperature $T_a$, we must integrate for time $t = (2/\Delta f)(T_s/sT_a)^2$. For $\Delta f = 1x10^8$ Hz (to get some resolution of the source spectrum), $T_s = 250$ K, s = 0.2 (5σ), we can reach a flux level of $3x10^{-20}$ $Wm^{-2}$ in an integration time of 18,400 s or 5 hours. This means that fairly lengthy integration times are going to be required and a focal plane array will have a real advantage if the system allows simultaneous observation of multiple sources of interest within the telescope's field of view.

The 210 to 310 GHz range, defined largely by a window of very high atmospheric transparency, is very effective for this type of study. As indicated in Fig. 1, there will almost always be multiple CO transitions observable. This is important for eliminating ambiguity in a redshift determined from a single line. It is advantageous to cover a wide a frequency range in order to maximize redshift coverage and the number of transitions observable simultaneously.

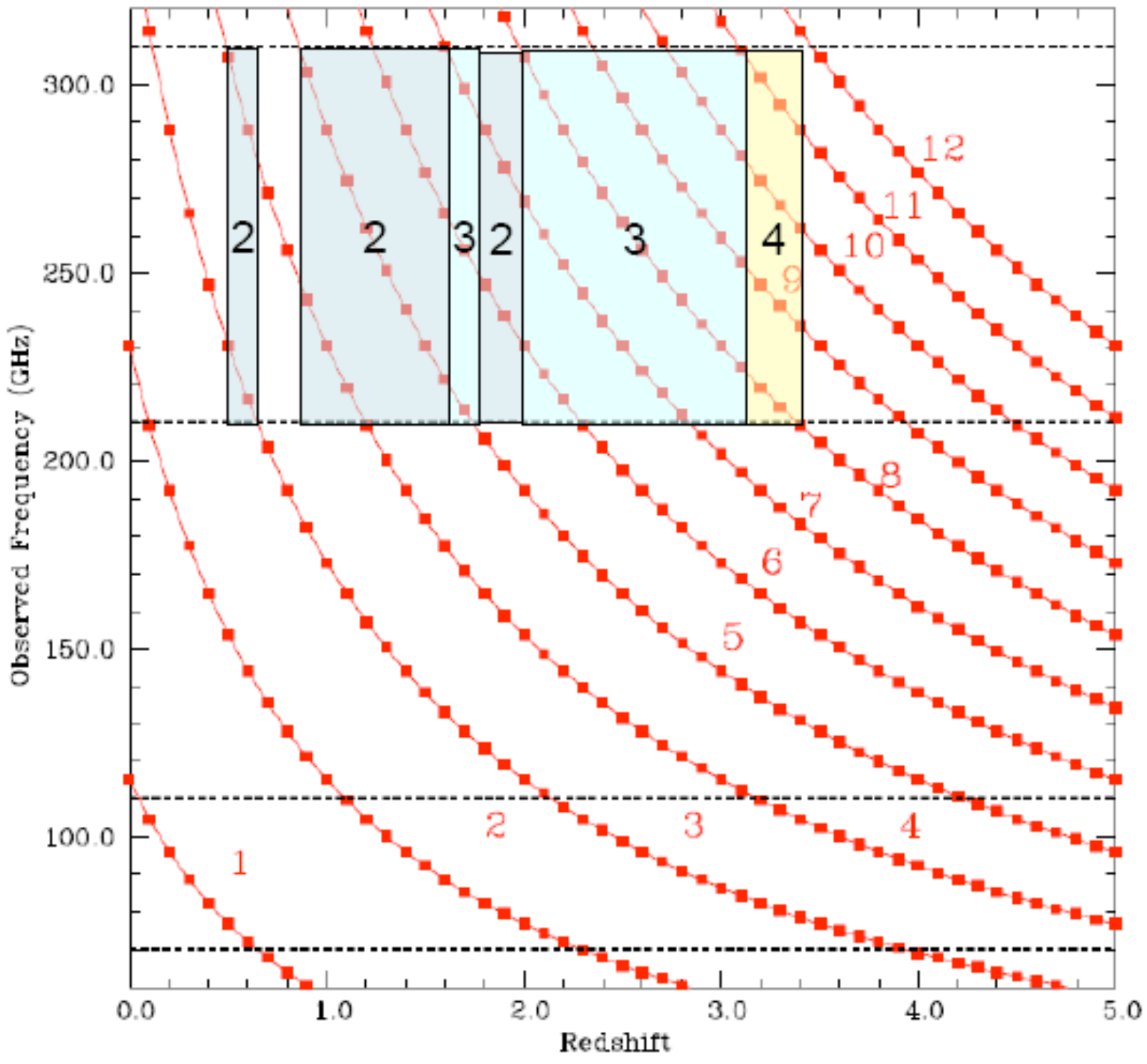

Fig. 1. Schematic showing observed frequency of various CO rotational transitions as function of source redshift. The numbers in the shaded boxes indicate the number of different transitions that can be observed with a system covering 210 to 310 GHz. For low redshifts and for the region around z = 0.8, only a single transition is observable, but in all other redshift ranges, more than one transition can be observed.

## 2. CCAT FOV AND NUMBER OF SOURCES

The nominal field of view (FOV) of the CCAT telescope is specified to be 20’ with a very modest degradation of the aperture efficiency (or effective area) at the periphery of the field (Cortés-Medellín & Herter 2006). This is an area of 0.11 square degrees. If we have 330 sources per square degree brighter than $3 \times 10^{-20}$ $Wm^{-2}$, there will be (on average) 36 such sources per CCAT FOV. This suggests that this is the order of magnitude of the number of receivers that can be profitably employed. However, these sources can be distributed anywhere within the FOV, which has a physical diameter equal to 87 cm. We thus have the significant challenge of how to couple the inputs of the some tens of receivers to an equal number of spots in the focal plane distributed over this area.

## 3. CONCEPTUAL DESIGN OF MOS WITH REFLECTIVE OPTICS

The concept for the MOS using reflective optics is derived from work on systems to position the fibers in a multi-object spectrograph operating at optical wavelengths (Moore *et al*. 2003, Moore & McGrath 2004, Moore *et al*. 2006, Schlegel & Ghiorso, 2008). The idea is to break up the focal plane into a number of “patrol regions”. Each region can send the signal from one spot in the focal plane located within it to a single receiver. We are assuming that the positions of the sources are known from previous observations, *e.g*. broadband imaging surveys utilizing large format array detectors. There is only modest overlap between patrol regions, which means that beyond a certain source density, not all sources can be simultaneously observed. While certainly not the only approach, a reasonable starting point is to spread the receivers available across the focal plane in a hexagonal close-packed array. This is based on the assumption that the sources are uniformly distributed on the sky.

In our design, a series of four reflectors couples each receiver feed horn to the focal plane, so we denote our optical system the "Mirror Multi-Object Spectrometer" or Mirror MOS. The mirrors are connected by pivoting arms, which allow the patrol region to be completely covered, as shown in Fig. 2.

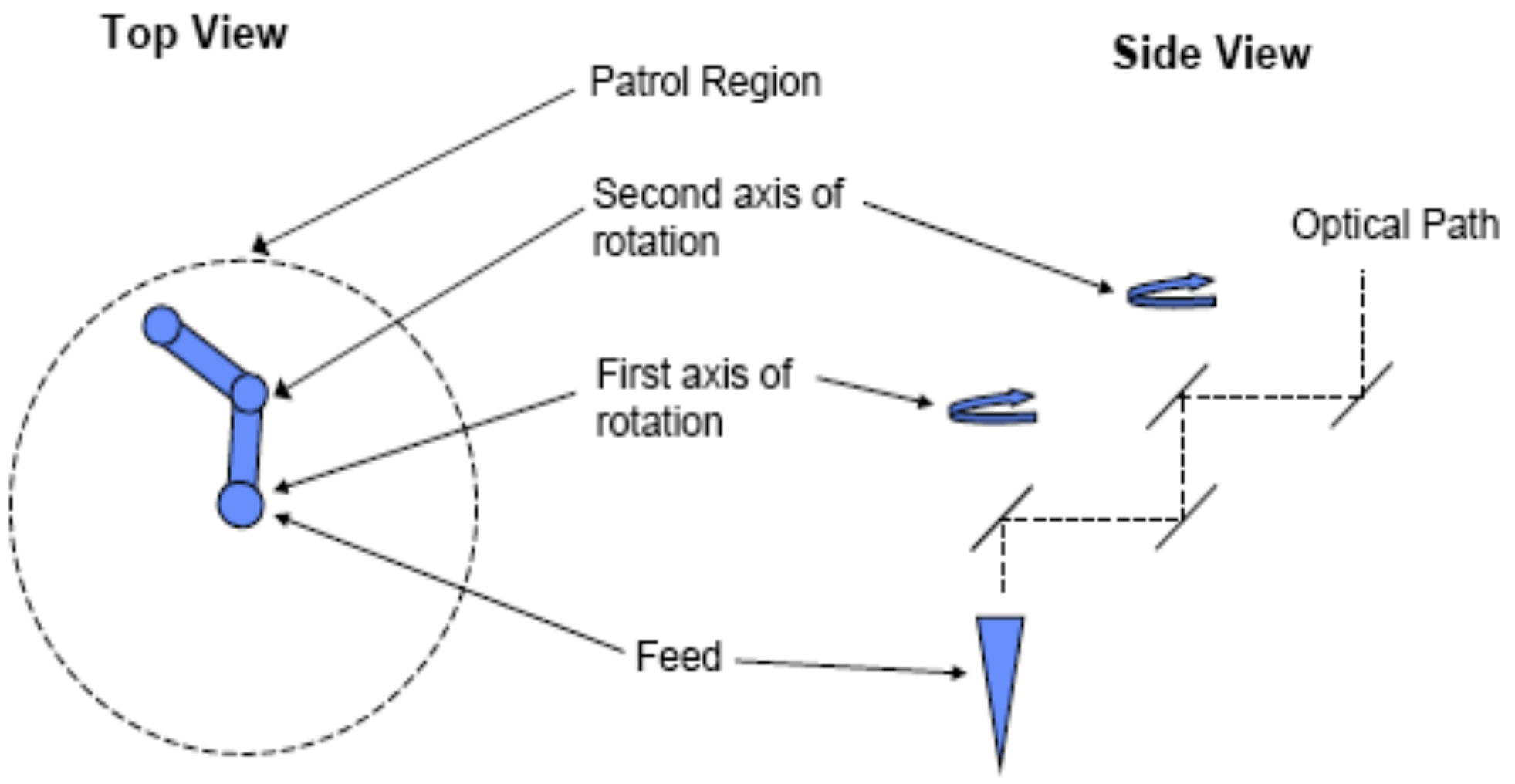

Fig. 2. Schematic diagram of a single receiver feed and the two arm motions that couple the beam from the feed to a given position in the focal plane.

The rotation about the first axis of rotation swings all four mirrors in a circle centered on the axis of the feed. The second axis of rotation swings the two mirrors furthest from the feed in another circle centered at the end of the arm defined by the two mirrors nearest the feed horn. It is evident that this arrangement can couple the feed to any point within a circle of radius equal to the sum of the lengths (measured perpendicular to the feed axis) of the two arms. To ensure complete coverage of the focal plane, the spacing of the feeds, which we denote s, should be somewhat smaller than twice the sum of these arm lengths, which we call the patrol radius and denote p. Fig. 2 above shows the reflectors as planar mirrors, which can work, but curved surfaces allow a more compact system, as discussed further below.

## 4. DESIGN EXAMPLE

The above system can be applied to arrays of arbitrary number of receivers, and can have numerous variations in terms of arm lengths (equal and unequal), as well as different geometries for the mirrors. In order to show how this system might be useful for a practical system to study SMGs and be consistent with the source densities discussed above, we choose some plausible parameters to show the viability of this approach.

1. There are certain “magic” numbers of receivers for hexagonal arrays, in terms of covering a circular region in the focal plane. These have 7, 19, 43, … receivers. We adopt a straw man array geometry having 19 receivers.

2. We take the spacing of the receivers s and the aperture plane diameter D to be related by s = D/5. Thus, the full 20’ aperture plane defining D = 87.3 cm requires s = 17.5 cm, but we adopt s = 15 cm in order not to overemphasize coverage of the edge of the focal plane where the efficiency will be lower. Another factor in this regard is the field curvature; the present system is actually very well adapted to correct for this, which can be accomplished by changing the location of the mirrors along an axis parallel to the feed horn axes. However, we will assume for the present time that the 75 cm diameter region can be covered without any refocusing. This results in the configuration shown in Fig. 3 below.

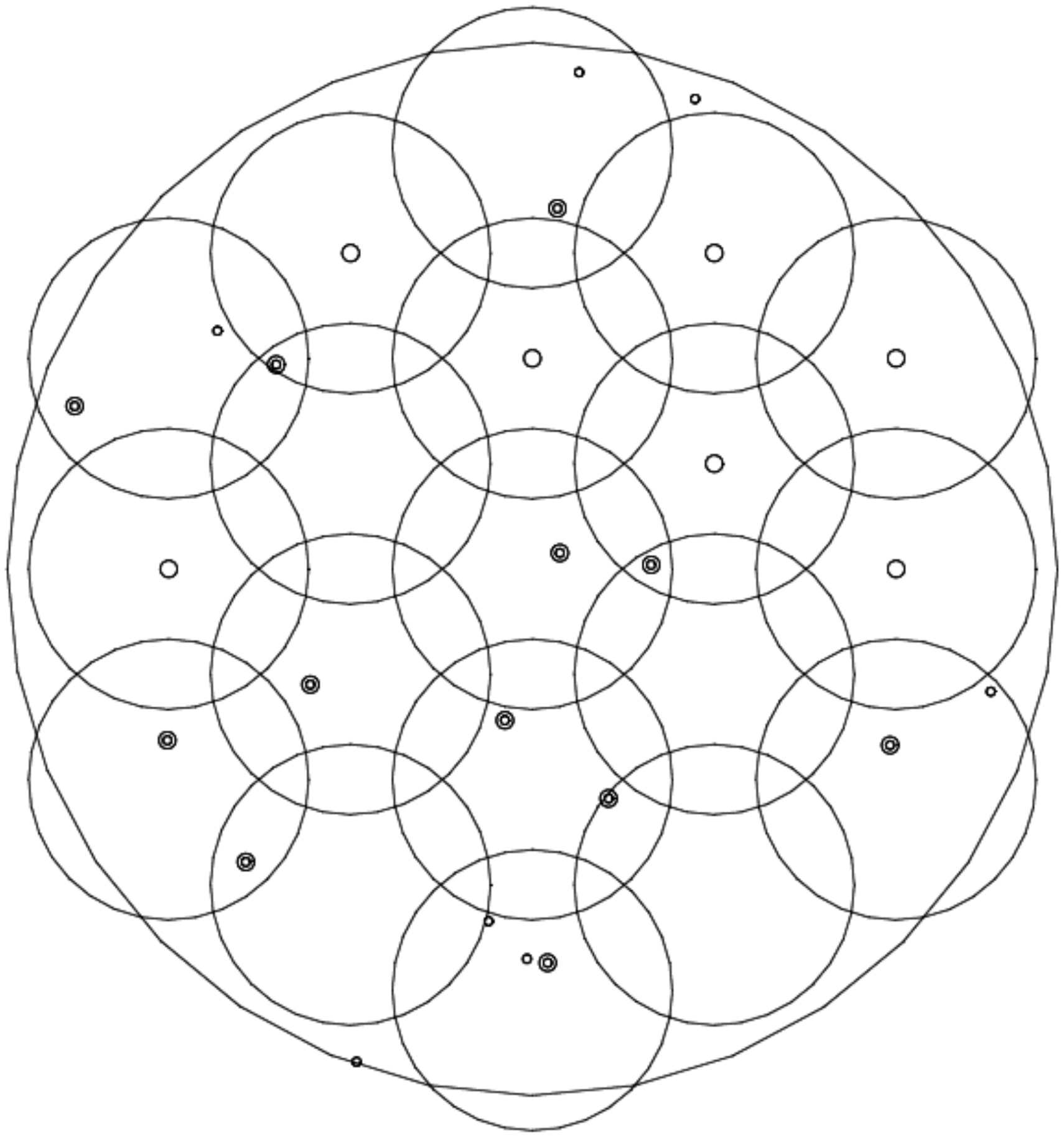

Fig. 3. Geometry of 19-receiver MMOS covering the CCAT focal plane. The largest circle shows the selected 75 cm diameter region of the focal plane. The separation of the receivers is s = 15 cm and the patrol radius p = 10 cm, indicated by the 19 circles of next-smaller size. The smallest circles represent the locations of 19 randomly distributed astronomical sources. The slightly larger circles represent the locations of the beam centers. In the cases where a beam is not assigned to a source, it is indicated as remaining in the center of the patrol region. In the present realization with 19 sources, 12 beams are assigned to sources, 7 beams are not assigned to sources and are thus unused, and 7 sources are not observed.

3. To avoid spillover loss, the reflective mirrors have minimum diameter, which is a significant system constraint at longer wavelengths. This results in a minimum separation of two pixels in the focal plane. This avoidance distance, denoted a, is set to 5 cm for the present example based on discussion of Gaussian beam propagation which follows.

## 5. GAUSSIAN BEAM PROPAGATION ANALYSIS

High quality feeds at millimeter and submillimeter wavelengths produce nearly fundamental mode Gaussian radiation patterns and their radiation can thus be effectively modeled by a single Gaussian beam, as discussed at length in Goldsmith (1998). The first step in this process is to determine the required beam waist in the telescope focal plane, which is given by the formula

$$w_0 = 0.22\ [T_E(dB)]^{0.5}\ (f/D)\ \lambda. \qquad (1)$$

In this expression, $T_E$(dB) is the edge taper of the illumination of the primary reflector in dB, f/D is the focal ratio of the telescope system where the receiver is located, and $\lambda$ is the wavelength. The distribution of power density in the beam as a function of the distance from the axis of propagation r, is given by

$$P(r)/P(0) = \exp[-2(r/w)^2], \quad (2)$$

where w is the beam radius. This quantity is a function of distance z along the axis of propagation from the beam waist, with the variation given by

$$w(z) = w_0[1 + (z/z_c)^2]^{0.5} , \quad (3)$$

where $z_c$ is the confocal distance, given by

$$z_c = \pi w_0^2/\lambda . \quad (4)$$

At a distance z equal to the confocal distance from the waist, the beam radius is a factor √2 larger than its minimum value, the beam waist radius $w_0$, which occurs at the beam waist.

For maximum aperture efficiency, the edge taper of a telescope should be close to 11 dB. The Nasmyth focus of the CCAT telescope is specified to have a focal ratio of 8. At a wavelength of 0.12 cm (frequency 250 GHz), which we adopt for this study, Eq. 1 results in a waist radius of 0.7 cm. To achieve low spillover requires a mirror diameter equal to 4w. From Eq. 2 this results in approximately 0.001 of the power spilling past the mirror, but with multiple mirrors in the system and non-Gaussian contributions to the beam distribution, this is a reasonable criterion. Thus, even near the waist, a minimum mirror diameter is 2.8 cm.

With the patrol radius equal to 10 cm, the path length between each of the mirrors in Fig. 2, as well as from the feed to the first mirror, is 5 cm. The input plane to the whole system is then approximately 20 cm from the feed horn, by which point the beam radius is ~ 1.3 cm and the beam (and mirror) diameter is 5.2 cm. Thus, in a system with all flat mirrors, the relevant mirror diameter is just over 5 cm. This makes the geometry shown in Fig. 2 feasible, with mirrors of this projected size and path lengths between the elements also equal to ~ 5 cm. The advantage of this system is that there is negligible beam distortion and cross polarization. The only significant loss is the ohmic loss of the mirrors and the spillover.

## 6. SYSTEM EFFICIENCY

We can calculate an "allocation fraction" for the system in the following manner. We assume that there is a specified density of sources which are considered targets for observation, and that they are randomly distributed in the telescope focal plane. For a given receiver separation, patrol radius, and avoidance distance, we can then calculate the fraction of sources which can be allocated to a target. The results of such simulations are shown in Fig. 4.

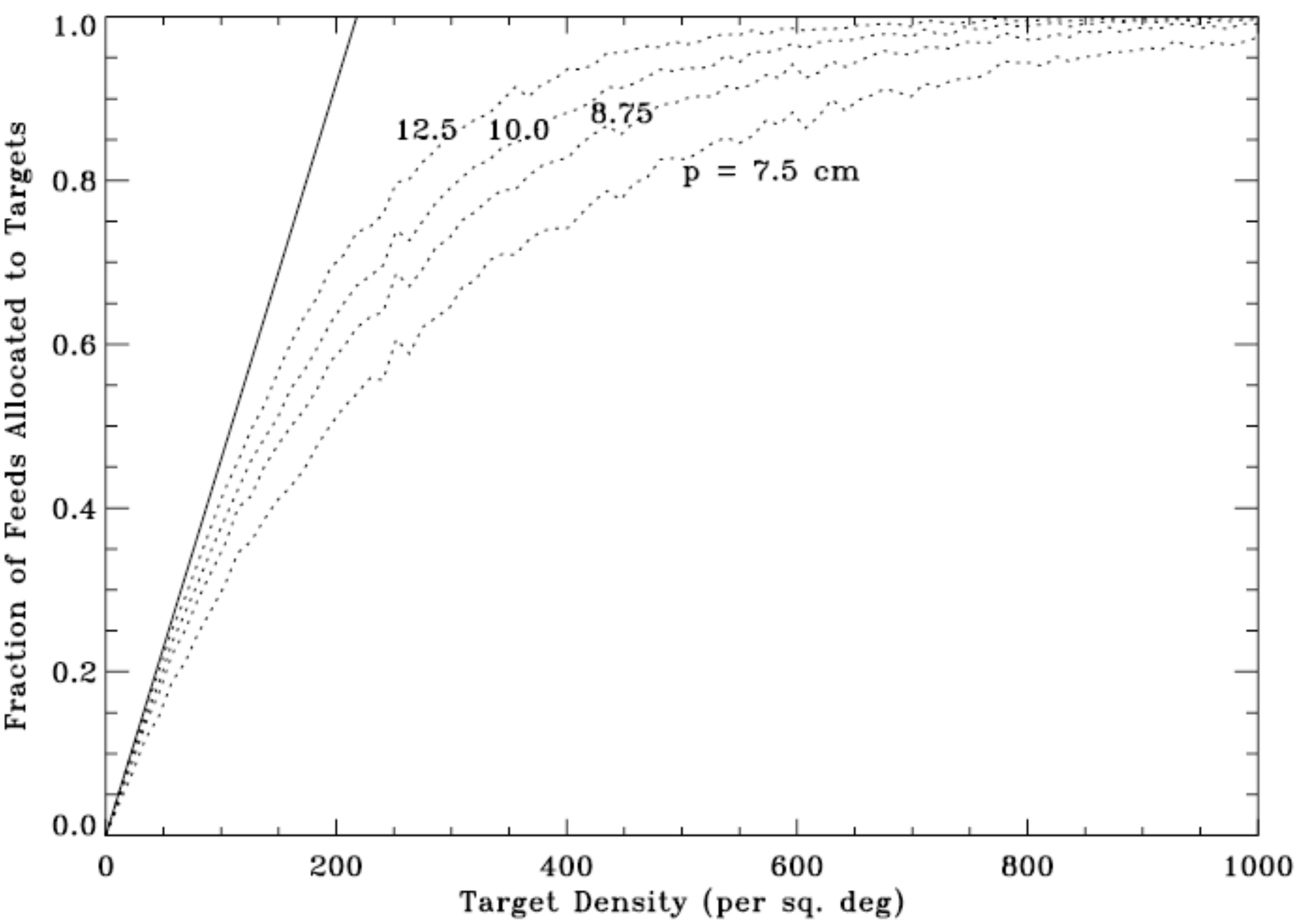

Fig. 4. Allocation fraction of mirror MOS. The solid curve is for an ideal 19-element system, which can place feeds anywhere in the 75 cm diameter field of view. The four dotted lines are for simulations with patrol radii of 12.5, 10, 8.75, and 7.5 cm, and an avoidance distance a = 5 cm. For very high target densities, the fraction of feeds allocated approaches unity and the patrol radius is not important because there will almost always be a target within reach of every feed. Each of the dotted curves is the average result of 100 realizations.

For 330 sources per square degree brighter than $3 \times 10^{-20}$ $Wm^{-2}$, the patrol radius does make a difference, and a larger radius is helpful, but the problem for $p \gg s$ is that there is an increased likelihood of "collisions" between the beams. For a patrol radius $p = 10$ cm, approximately 85% of the feeds will be allocated.

## 7. SYSTEM EMPLOYING REIMAGING OPTICS

As discussed above, a reflective optics MMOS system operating at a wavelength of 0.12 cm can be made using only plane mirrors. This is certainly the simplest system, but the inclusion of reimaging optics has some potential advantages that should be considered. Specifically, we wish to replace one or more flat mirrors with reimaging optical elements. These could be ellipsoidal reflectors, or off-axis paraboloids. Some relevant points are that

1. reimaging optics will limit the maximum beam radius within the system, which will result in smaller elements and thus smaller avoidance distances. Reimaging optics can also be used to yield lower spillover and thus higher overall efficiency,

2. reimaging optics will change the behavior of the effective beam waist radius of the system, which will influence the system performance when a wide frequency range is considered, and that

3. reimaging optics can produce cross polarization and beam distortion, which should be evaluated for any specific configuration.

Reimaging optics and Gaussian beams are discussed at length by *e.g.* Goldsmith (1998). For our purposes we consider only the very simplest situation consisting of an element with a specified focal length f. For a Gaussian beam having input waist radius $w_o$ in at distance f from the focusing element, the output beam waist will also be at distance f, and the output beam waist radius $w_{o\ out}$ is given by

$$w_{o\ out} = \lambda f / \pi w_{o\ in}. \tag{5}$$

We can consider substituting a focusing mirror for one of the flat mirrors, as shown schematically in Fig. 5.

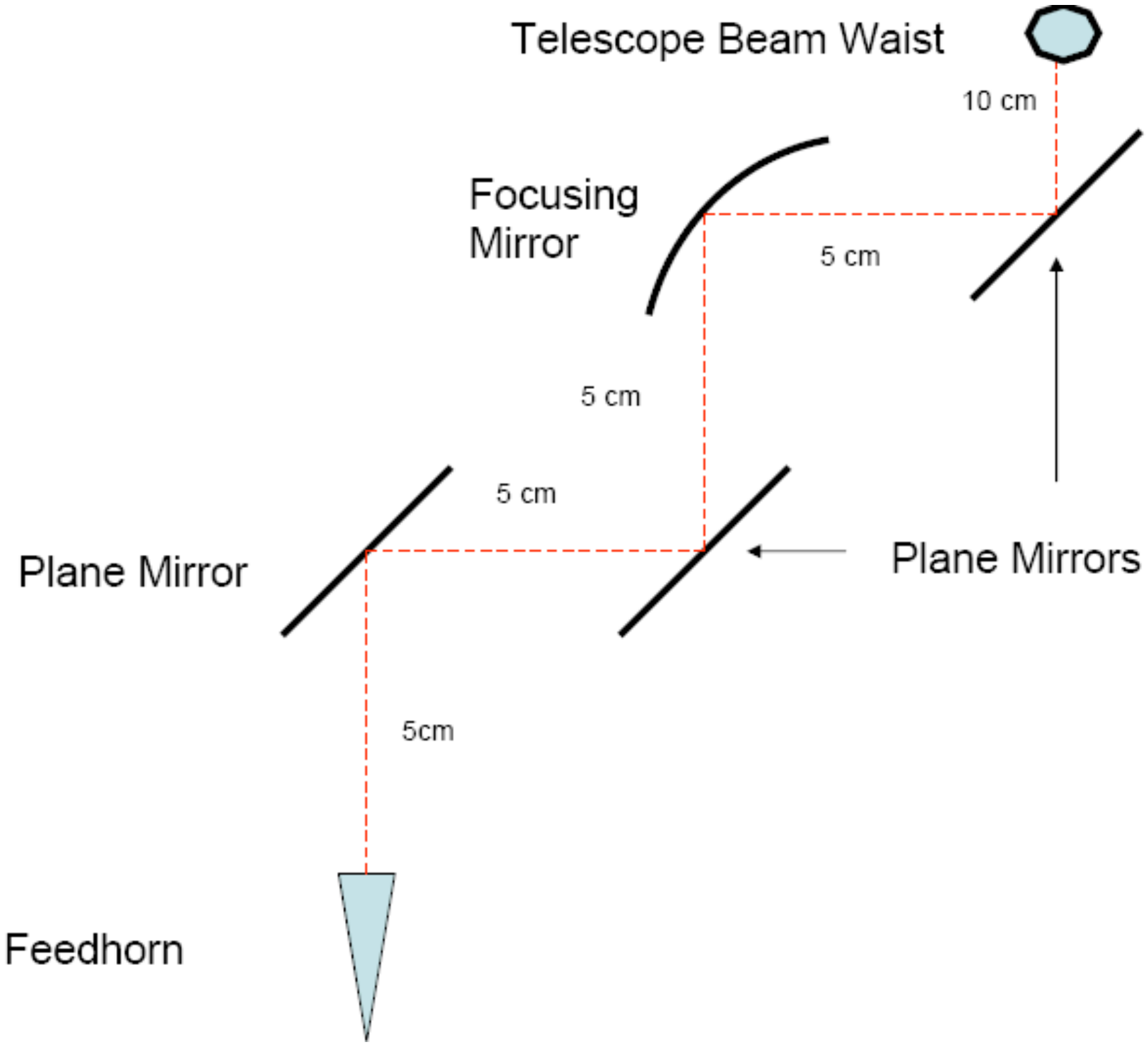

Fig. 5. Schematic of reflective MMOS input optics employing a single focusing mirror and three plane mirrors. In the design considered here, the mirror focal length is f = 15 cm, and the telescope beam waist and the feedhorn beam waist are both located at this distance from the focusing mirror.

In this case, we assume that the distance from the telescope beam waist to the focusing mirror is 15 cm (of which 5 cm is the second moving arm). The focal length of the off-axis ellipsoid is 15 cm, meaning that the input telescope beam waist ($w_0$ = 0.7 cm) is transformed to an output beam waist having $w_0$ = 0.82 cm. The choice of distances and focal length mean that this output beam waist is located at the aperture of the feedhorn. The beam radius at the ellipsoidal mirror is 1.1 cm, so the mirror diameter could be as small as 4.5 cm, while the other mirrors could be even smaller. Thus, the reimaging does allow a modest reduction in the mirror size and avoidance distance, and this process could be pushed even further by using focusing mirrors in place of the remaining plane mirrors.

One effect of this reimaging is that the feedhorn beam waist is somewhat larger (0.82 cm compared to 0.7 cm), meaning that the feedhorn diameter (typically ~ $3w_0$) is also slightly larger. However, there is an important advantage. An aperture-limited feedhorn has a beam waist that is essentially independent of wavelength, so that the beam size varies linearly with the wavelength. This is not desirable for broadband systems, because the edge taper of the primary reflector produced by direct coupling to such a feedhorn is too weak at low frequencies and too strong at high frequencies, both resulting in a reduction in efficiency.

With the present single reimaging element, the 0.82 cm feedhorn waist is transformed into a 0.7 cm telescope beam waist (using eq. 5), $w_{0\,tel} = \lambda f/\pi w_{o\,in} = 0.7$ cm ($\lambda$/0.12 cm). The Gaussian beam diverging from the telescope beam waist is, once in the far field ($z >> z_c \approx 12.5$ cm), independent of wavelength. This gives uniformly high efficiency over a very large bandwidth. This desirable behavior can be preserved if there are an odd number of such reimaging stages between the feedhorn and the telescope beam waist.

It is possible to design more complex reimaging systems, and this can effectively be done by adding a Gaussian beam telescope, consisting of a pair of focusing elements separated by the sum of their focal lengths. Such a device produces a wavelength-independent magnification of the Gaussian beam waist, by a ratio M = ratio of element focal lengths. Gaussian beam telescopes could be used to pack the receivers more closely together in the center of the field, for example, which would result in a smaller dewar. The number of design possibilities is very large, and we do not explore these further here.

## 8. VARIOUS PERFORMANCE CONSIDERATIONS

### 8.1 Beam Distortion and Cross Polarization

The use of an off-axis focusing element does produce some beam distortion and cross polarization, as discussed in Goldsmith (1998). The focusing element here is relatively "slow" and the fraction of power lost from the fundamental Gaussian beam mode is less than 0.001. The fraction of power converted to the orthogonal polarization state is twice as large, but is still only about 0.0014. It does not seem that in the present application either slight beam distortion or cross polarization is a significant issue, so that the use of multiple reimaging elements seems entirely feasible. It is also possible to design the various elements such that their undesirable behaviors cancel to some extent, which would further reduce their deleterious effect.

## 8.2 Mirror Emissivity

Metal reflectors in short mm/submm range have finite emissivity. In detail, this depends on the angle of incidence and polarization state, but to get a feel for the situation, we can use the measured values from Bock et al. (1995) who report emissivity $\varepsilon = 3.2\text{x}10^{-3}$ at $\lambda = 0.12$ cm. This is only moderately greater than the theoretical value for Al, $1.85\text{x}10^{-3}$ at $45^{o}$ angle of incidence. For our 4 mirrors, the total emissivity would thus be approximately $1.3\text{x}10^{-2}$. This is almost certainly less than the loss in a dewar window, for example, and even if the mirrors are at ambient temperature, adds only 4 K to the noise temperature of a coherent system. This is also considerably less than the effective temperature contributed by the atmosphere at these wavelengths, so that the mirror emission would be a relatively small contribution to the total power collected by a coherent or an incoherent detector system.

## 8.3 Beam Truncation and Spillover

If we maintain a mirror diameter equal to 4w (measured perpendicular to the axis of propagation), the fraction of power spilling past a circular mirror is 0.0003. In general, when one has a sequence of mirrors such as is the case here, the total loss to the fundamental Gaussian mode is not necessarily the sum of the values of each individual beam truncation, due to diffraction which can produce phase ripples and other harmful effects which do not add together in any simple fashion. If we do maintain the truncation level for each element at the $< 0.001$ level, such effects should be minor. For the present calculation, we can simply add the spillover losses for the 4 elements to get a combined loss of 0.0012, noting that a full diffraction analysis should be carried out.

## 9. SOME MECHANICAL DETAILS

We show a schematic of the mirror MOS in Fig. 6. The basic mechanical concept includes two "stacked" periscopes each of which pivots on an open bearing. Periscope 1 has to be supported below its bearing by a bracket which does not block the beam, while periscope 2 rotates relative to periscope 1. Each periscope has a drive motor and encoder. The electrical connections for periscope 2 drive have to be brought to the fixed mounting that holds the feedhorn and receivers. This would require slip rings, or else the angular rotation would have to be limited to avoid breaking wires. In the latter approach, it is important that a minimum of 360 degrees rotation capability be maintained so that the entire circular patrol region is accessible.

The following gives an idea of the tolerances for the periscope system. The position error of the beam in the focal plane depends on the two angles, but a simple, representative case is that in which the two periscopes are parallel. In this case, the angular position error is equal to the patrol radius multiplied by the sum of the angular errors of the two periscopes. A positional error corresponding to a tenth of a beam waist radius corresponds to 0.07 cm at 0.12 cm wavelength, so with a patrol radius of 10 cm, the sum of the angular errors must be < 0.007 radian or $0.4^{\circ}$. The individual periscope angular errors must then be $< 0.2^{\circ}$ or 0.0035 radian. This corresponds to an angular resolution of 1 part in 300. An encoder with 10 bits resolution should satisfy this requirement, and 12 bits would leave a good margin.

For a focal ratio of 8, the beam divergence angle is ~ 7°, so that the overall tilt errors in the dual periscope system must be held to a small fraction of this value to assure optimum illumination of the primary. This can be analyzed more quantitatively for a given illumination pattern and specified efficiency loss, but a maximum angular error of approximately 0.5° appears to be a reasonable value. The requirements are dramatically less demanding than those for optically positioning systems, and should not be difficult to achieve without resort to any complex or expensive bearings or mechanical structure.

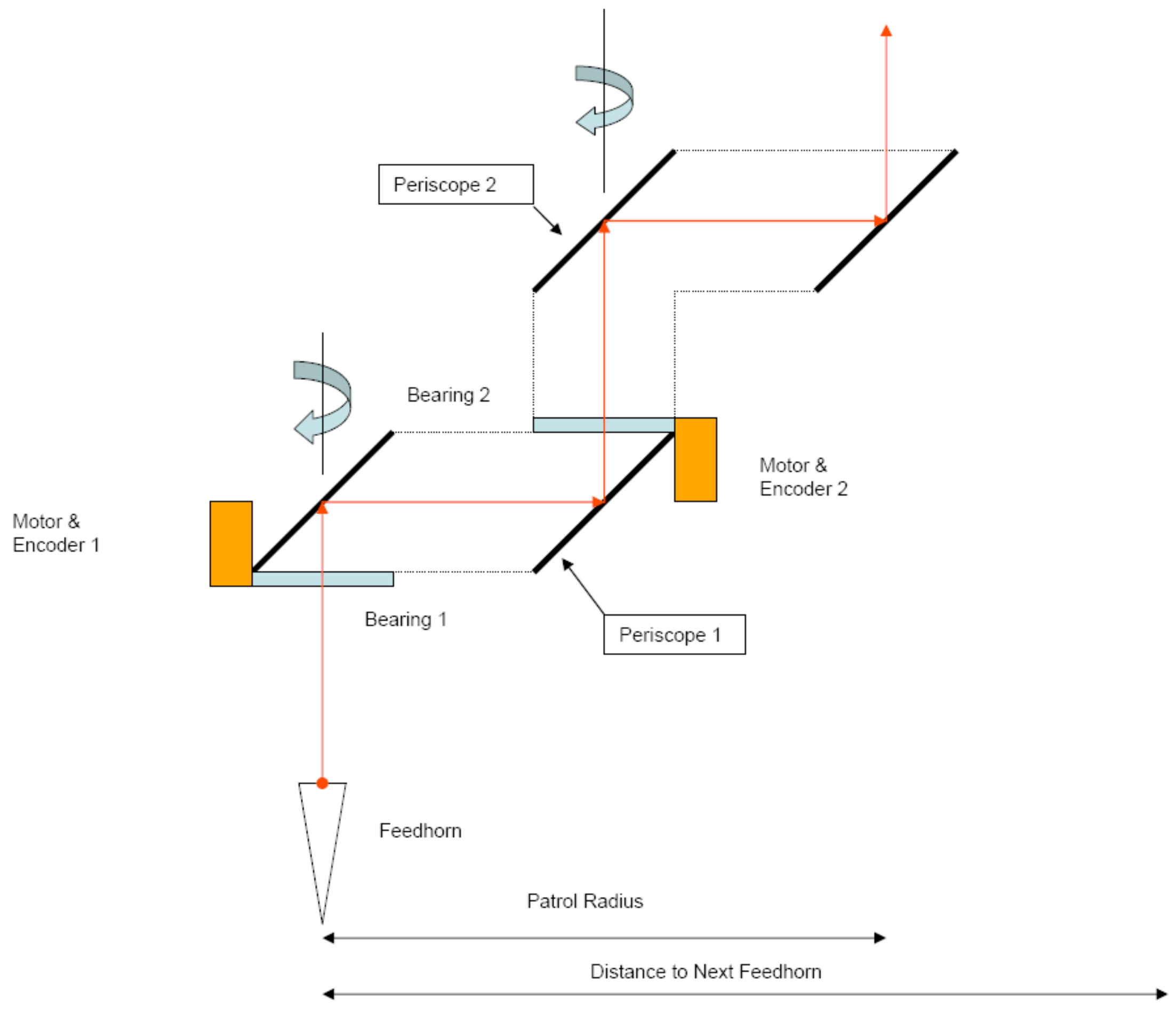

Fig. 6. Schematic of the mechanical arrangement of the dual periscope mirror MOS system. The axes of rotation of each periscope, consisting of a pair of mirrors, are indicated, along with support bearings, motors, and encoders. The use of focusing rather than planar optics does not affect this basic design.

## 10. SUMMARY

We have described the mirror MOS system for coupling the inputs of a multiobject spectrometer to the focal plane of a large submillimeter telescope using only reflective optics. By dividing up the focal plane into patrol regions, a system with good allocation efficiency of feeds for anticipated source densities can be realized. This approach has not been optimized, but we conclude that while it can be realized with only planar mirrors, it can be made more compact, with better coverage of the focal plane, using one or more refocusing elements. We have analyzed such a system in terms of fundamental mode Gaussian beam propagation, and find that the beam distortion and cross polarization levels are very low. This approach also allows us to transform a frequency-independent high quality beam waist, as is produced by a low aperture phase error corrugated feed, to a telescope beam waist which produces wavelength-independent illumination of the telescope. The estimated levels of loss to the fundamental Gaussian mode due to beam distortion, cross polarization, and spillover are each at approximately the 0.001 level, while ohmic loss is a factor of 10 larger, together suggesting that this the reflective Gaussian beam multi-object spectrometer will be more flexible and have lower loss than other techniques for coupling feedhorns to arbitrary positions in the focal plane, such as flexible waveguides. The mirror MOS should be an extremely effective approach for simultaneous observation of multiple sources with CCAT and other submillimeter telescopes.

We thank Andrew Blain for helpful suggestions regarding source densities and observing strategies, and an anonymous referee for useful comments. The research described in this paper was carried out at the Jet Propulsion Laboratory, California Institute of Technology, under a contract with the National Aeronautics and Space Administration.